\documentclass[aps,pra,amsmath,amssymb, reprint,superscriptaddress, floatfix]{revtex4-2}
\newcommand{\bol}[1]{\boldsymbol{#1}}
\usepackage{physics}
\usepackage{calc}
\usepackage{mathtools}
\newcommand{\cmmnt}[1]{}
\usepackage{tikz}
\usepackage{enumitem}
\usepackage{balance}
\usepackage{longtable}
\usepackage[pdftex,pdfpagelabels,bookmarks,hyperindex,hyperfigures,colorlinks]{hyperref}

\begin{document}


\title{A Quantum Walk Inspired Qubit Lattice Algorithm for Simulating Electromagnetic Wave Propagation and Scattering in Conservative and Dissipative Magnetized Plasmas}


\author{Efstratios Koukoutsis}
\email{stkoukoutsis@mail.ntua.gr}
\author{Kyriakos Hizanidis}
\affiliation{School of Electrical and Computer Engineering, National Technical University of Athens, Zographou 15780, Greece}
\author{George Vahala}
\affiliation{Department of Physics, William \& Mary, Williamsburg, Virginia 23187, USA}
\author{Christos Tsironis}
\affiliation{School of Electrical and Computer Engineering, National Technical University of Athens, Zographou 15780, Greece}
\author{Abhay K. Ram}
\affiliation{Plasma Science and Fusion Center, Massachusetts Institute of Technology, Cambridge,
Massachusetts 02139, USA}
\author{Min Soe}
\affiliation{Department of Mathematics and Physical Sciences, Rogers State University, Claremore, Oklahoma 74017, USA}
\author{Linda Vahala}
\affiliation{Department of Electrical and Computer Engineering, Old Dominion University, Norfolk, Virginia 23529, USA}


\date{\today}

\begin{abstract}
Based on the Dirac representation of Maxwell equations we present an explicit, discrete space-time, quantum walk-inspired algorithm suitable for simulating the electromagnetic wave propagation and scattering from inhomogeneities within magnetized plasmas. The quantum walk is implemented on a lattice with an internal space of $n_q=4$--qubits, used to encode the classical field amplitudes. Unitary rotation gates operate within this internal space to generate the non-trivial dynamics of the free plasma-Dirac equation. To incorporate the contributions from the cyclotron and plasma density terms--manifesting as inhomogeneous potential terms--in the plasma-Dirac equation, the walk process is complemented with unitary potential operators.
This leads to a unitary qubit lattice sequence that recovers the plasma-Dirac equation under a second-order accurate discretization scheme. The proposed algorithm is explicit and demonstrates, in the worst case, a polynomial quantum advantage compared to the Finite Difference Time Domain (FDTD) classical method in terms of resource requirements and error complexity. In addition, we extend the algorithm to include dissipative effects by introducing a phenomenological collision frequency between plasma species. Then, a post-selective time-marching implementation scheme is delineated, featuring a non-vanishing overall success probability and, subsequently, eliminating the need for amplitude amplification of the output state while preserving the quantum advantage.
\end{abstract}


\maketitle

\section{Introduction}\label{sec:1}
Quantum walks (QW) being the quantum counterparts of classical random walks \cite{Aharonov_1993,Ambainis_2001} play important role in quantum computing and particularly in quantum simulation \cite{Childs_2003, Childs_2010,Berry_2012,Berry_2015}. Also, they have been  established as universal model of quantum computation \cite{Childs_2009}. In particular, discrete-time quantum walks (DTQW) on regular lattices can give rise to wave equations for relativistic particles in the continuum limit \cite{Gourdeau_2017,Mlodinow_2018,Nzongani_2024} enabling an efficient simulation process. In the standard notation of DTQW, the dynamics of a particle are described by a walking exterior space $\mathcal{H}_S$ and an interior $2$-dimensional Hilbert space $\mathcal{H}_{C}$ dubbed as the coin/spin space in which different unitary coin operators $\hat{C}$ generating various non-trivial dynamics. The walking process is applied between the vertexes of the lattice though the streaming unitary operator $\hat{S}$ acting on the $\{\ket{p}\}\in\mathcal{H}_S$ register in respect of the spin register,
\begin{equation}\label{1d quantum walk}
\hat{S}=\ket{0}\bra{0}\otimes\ket{p+1}\bra{p}+\ket{1}\bra{1}\otimes\ket{p-1}\bra{p}.
\end{equation}
Thus, the evolution of the state $\ket{\psi(t)}\in\mathcal{H}=\mathcal{H}_C\otimes\mathcal{H}_S$,
\begin{equation}\label{state}
\ket{\psi(t)}=\sum_{p=0}^{2^{n_p} -1}(\psi_0(t)\ket{0}+\psi_1(t)\ket{1})\otimes\ket{p}
\end{equation}
from time $t$ to $t+\Delta t$ is 
\begin{equation}\label{qw step evolution}
\ket{\psi(t+\Delta t}=\hat{S}(\hat{C}\otimes \hat{I})\ket{\psi(t)}.
\end{equation}
For a general $3$-dimensional collocated lattice, (resulting from the discretization of the configuration space $\mathcal{V}=[j_0, j_0+L_j]^3\subset\mathbb{R}^3$, comprised of $N_j$ nodes separated by $\delta_j=L_j/N_j$ for each axis $j=x,y,z$), the $\ket{p}$ state describes the position of the particle in each lattice node,
\begin{equation}\label{spatial register}
\ket{p}=\bigotimes_j\ket{p_j},\quad \ket{p_j}=\ket{j_0+p_j\delta_j},\quad n_p=\sum_j \log_2{N_j}.
\end{equation}
In Eq.\eqref{spatial register}, $n_p$ is the number of qubits characterizing the $\ket{p}$ state. In contrast with random walks the evolution in Eq.\eqref{qw step evolution} is unitary, hence reversible.

On increasing the dimension of the spin space $\mathcal{H}_C$ into $d$-dimensions, we can extend  QW algorithms to multi-dimensional and multi-particle quantum secular automata \cite{Birula2_1994,Boghosian_1998,Yepez3_2002}, quantum lattice Boltzmann \cite{Succi_2015} and, eventually, qubit lattice algorithms (QLA) \cite{Yepez_2002,Vahala3_2023}. In particular, QLA’s backbone for the quantum representation of Maxwell equations in a general passive electromagnetic medium is based on the massless Direac-type equation. This has then permitted quantum extensions to handle Maxwell equations in complex media \cite{Vahala_2020,Koukoutsis_2023,Vahala_2023,Vahala2_2023,Koukoutsis2_2023,Koukoutsis_2024}. The advantage of the aforementioned QW-inspired algorithms lie on the fact that the walking process can be implemented efficiently in the lattice as it will be showcased later.

Classical Maxwell equations have  recently emerged as a compelling set of differential equations for applying quantum algorithms  \cite{Sinha_2010,Costa_2019,Zhang_2021,Novak_2023,Novikau_2023,Jin_2024,Nguyen_2024}, primarily due to their $\mathbf{(i)}$ inherent linearity within the linear response framework and $\mathbf{(ii)}$ broad applicability to various physical problems. Albeit the significant contributions of the previous studies, most have focused on simplified models of wave propagation and scattering in either homogeneous or inhomogeneous scalar media. In addition the  quantum implementation is based on unitary oracle operations, prohibiting an explicit implementation on quantum hardware. Such limitations mitigate the impact and the capabilities that these algorithms can have in realistic applications where the electromagnetic media response is anisotropic, inhomogeneous and potentially complex meaning that it can be dispersion and dissipation in the medium's response.

To this direction, based on the quantum representation of Maxwell equations in a cold magnetized plasma \cite{Koukoutsis2_2023}, by exploiting the Pauli structure of the generator of dynamics we present an explicit qubit lattice algorithm quantum encoded as a DTQW for simulation of electromagnetic wave propagation and scattering in inhomogeneous magnetized plasmas. The algorithm is explicit in terms of the required quantum resources and gate scaling as well as it exhibits a potential exponential quantum advantage compared to the contemporary and widely used Finite Difference Time Domain (FDTD) computational method \cite{Yee_1966,Schneider_1993} for studying electromagnetic wave scattering in plasmas \cite{Lee_1999,Tsironis_2023}.

The paper is organized as follows. Section \ref{sec:2.1} outlines the theoretical reformulation of Maxwell equations for a cold magnetized plasma as a quantum Dirac equation. In Sec.\ref{sec:2.2}, the details of encoding and discretization of the continuous plasma-Dirac system into qubit states are presented. Section \ref{sec:3} covers the algorithmic process, its explicit quantum circuit implementation, and the complexity scaling, demonstrating a quantum advatnage over the FDTD method. Section \ref{sec:4.1} introduces a phenomenological collisional dissipation process that breaks the unitary evolution of the conservative case. Then, Sec.\ref{sec:4.2} presents a post-selective quantum algorithm based on the LCU method  with an optimal overall success implementation probability.  As a result, the previously established quantum advantage is maintained.

\section{Quantum representation and encoding}\label{sec:2}
In this section we briefly revisit the theoretical construction of Maxwell equations in a cold magnetized plasma as a multi-spinor massless Dirac equation with a potential and expresses the electromagnetic state vector as a quantum state $\ket{\bol\psi}\in\mathcal{H}=\mathcal{H}_C\otimes\mathcal{H}_S$.

\subsection{Dirac representation of Maxwell equations in cold magnetized plasmas}\label{sec:2.1}
Cold magnetized plasmas are gyrotropic i.e. anisotropic dielectric media exhibiting temporal dispersion with a frequency dependent permittivity matrix $\Tilde{\epsilon}(\omega)$ in the frequency domain. Following the Stix notation \cite{Stix_1992},
\begin{equation}\label{Stix}
\Tilde{\epsilon}(\omega)=\begin{bmatrix}
S&-iD&0\\
iD&S&0\\
0&0&P
\end{bmatrix}
\end{equation}
with
\begin{align}\label{Stix elements}
S=&\epsilon_0\Big(1-\sum_{j=i,e}\frac{\omega^2_{pj}}{\omega^2-\omega_{cj}^2}\Big) \nonumber\\
D=&\epsilon_0\sum_{j=i,e}\frac{\omega_{cj}\omega^2_{pj}}{\omega(\omega^2-\omega_{cj}^2)}\\
P=&\epsilon_0\Big(1-\sum_{j=i,e}\frac{\omega^2_{pj}}{\omega^2}\Big) \nonumber.
\end{align}
The definition of the elements in Eq.\eqref{Stix elements} in the Stix permittivity tensor is taken for a two-species, ions (i) and electrons (e), plasma with inhomogeneous plasma frequency $\omega^2_{pj}(\bol{r})=\frac{n_j(\bol{r})q^2_j}{m_j\epsilon_0}$ where $n_j(\bf{r})$ is the $j^{th}$ species number density. The cyclotron frequency $\omega_{cj}=\frac{q_jB_{0}}{m_j}$ is defined in respect of a homogeneous magnetic field $B_0$  along the $z$ axis and $m_j$, $q_j$ are the mass and charge of the $j$-species respectively.

In the temporal domain,  the source free Maxwell equations in terms of the electromagnetic intensity $\bol d=(\bol D, \bol B)^T$ and electromagnetic fields $\bol u=(\bol E, \bol H)^T$ are compactly written as,
\begin{equation}\label{Maxwell in time}
i\pdv{\bol d}{t}=\hat{M}\bol u, \quad \div\bol d=0.
\end{equation}
The $\hat{M}$ operator in Eq.\eqref{Maxwell in time},
\begin{equation}\label{M}
\hat{M}=i\begin{bmatrix}
0_{3\times 3} & \curl{}\\
-\curl{} &0_{3\times 3}
\end{bmatrix}
\end{equation}
is a self adjoint operator $\hat{M}=\hat{M}^\dagger$ in the domain $\mathcal{D}(\hat{M})=L^2(\mathcal{V}\subset\mathbb{R}^3, \mathbb{C}^6)$ under the boundary condition $\bol n (\bol r)\times \bol E=0$ with $\bol n (\bol r)$ being the outward orthogonal vector in the boundary $\partial\mathcal{V}$. The divergence set of equations in Eq.\eqref{Maxwell in time} are treated as initial conditions.

Given the fact that the permittivity tensor $\Tilde{\epsilon}(\omega)$ is Hermitian there is conservation of a positive definite energy which is a suffice condition for the dynamics to be recast in an explicit quantum representation with Hermitian structure \cite{Mostafazadeh_2002}. As a result, following \cite{Koukoutsis2_2023}, transforming into the temporal domain we obtain an augmented version of Maxwell equations in a form of a massless  and multi-spinor Dirac equation with a potential $\hat{V}(\bol r)$,
\begin{equation}\label{plasma dirac equation}
i\pdv{\bol\psi}{t}=\Big[-c\hat{P}_{E,B}\otimes\hat{\bol\gamma}_{em}\cdot\hat{\bol p} +\hat
V(\bol r)\Big]\bol\psi.
\end{equation}
The $\bol\psi$ state in the plasma Dirac equation \eqref{plasma dirac equation} contain the pertinent electromagnetic fields and current densities,
\begin{equation}\label{state and conductivity}
\bol\psi=\begin{bmatrix}
\epsilon_0^{1/2}\bol E\\
\mu_0^{1/2}\bol H\\
\frac{1}{\epsilon_0^{1/2} \omega_{pi}}\bol J_{ci}\\
\frac{1}{\epsilon_0^{1/2} \omega_{pe}}\bol J_{ce}
\end{bmatrix},\quad \bol J_{cj}=\int_0^t\pdv{\hat{K}_j(t-\tau)}{t}\bol E(\bol r, \tau)d\,\tau,
\end{equation}
where $\hat{K}_j(t)$ is the susceptibility kernel contribution for each species in the temporal domain,
\begin{equation}\label{susceptibility kernel}
\begin{split}
\hat{K}(t)&=\epsilon_0\sum_{j=i,e}\begin{bmatrix}
\frac{\omega^2_{pj}}{\omega_{cj}}\sin{\omega_{cj}t}&\frac{\omega^2_{pj}}{\omega_{cj}}(\cos{\omega_{cj}t}-1) &0\\
\frac{\omega^2_{pj}}{\omega_{cj}}(1-\cos{\omega_{cj}t})&\frac{\omega^2_{pj}}{\omega_{cj}}\sin{\omega_{cj}t}&0\\
0&0&\omega^2_{pj}t
\end{bmatrix}\\
&=\hat{K}_i(t)+\hat{K}_e(t).
\end{split}
\end{equation}
The electromagnetic Dirac matrices $\hat{\bol\gamma}_{el}=(\hat{\gamma}_x,\hat{\gamma}_y, \hat{\gamma}_z)$ read,
\begin{equation}\label{gamma matrices}
\hat{\gamma}_i=\hat{\sigma}_y\otimes\hat{S}_i,\quad i=x,y,z,
\end{equation}
where $\hat{\sigma}_y$ is the Pauli $y$-matrix and $\hat{S}_i$ are the spin-1 matrices,
\begin{equation}\label{spin1 matrices}
\hat{S}_x=\begin{bmatrix}
0&0&0\\
0&0&-i\\
0&i&0
\end{bmatrix}\quad \hat{S_y}=\begin{bmatrix}
0&0&i\\
0&0&0\\
-i&0&0
\end{bmatrix}\quad \hat{S_z}=\begin{bmatrix}
0&-i&0\\
i&0&0\\
0&0&0
\end{bmatrix}.
\end{equation}
Finally, $c=(\epsilon_0\mu_0)^{-1/2}$ is the speed of light in the vacuum, $\hat{\bol p}=-i\nabla$ is the the quantum mechanical momentum operator, $\hat{P}_{E,H}=(\sigma_z+I_{2\times 2})/2$ is the projection operator in the subspace of the electromagnetic fields $\{\bol E, \bol H\}$  and the Hermitian potential operator $\hat{V}(\bol r)$ is \cite{Koukoutsis2_2023},
\begin{equation}\label{potential}
\hat{V}(\bol r)=\begin{bmatrix}
0_{3\times3}&0_{3\times3}&-i\omega_{pi}&-i\omega_{pe}\\
0_{3\times3}&0_{3\times3}&0_{3\times3}&0_{3\times3}\\
i\omega_{pi}&0_{3\times3}&\omega_{ci}\hat{S}_z&0_{3\times3}\\
i\omega_{pe}&0_{3\times3}&0_{3\times3}&\omega_{ce}\hat{S}_z
\end{bmatrix}.
\end{equation}

The positive definite conserved electromagnetic energy $E(t)$ reads,
\begin{equation}\label{energy}
\begin{split}
E(t)&=\int_\mathcal{V}\Big(\epsilon_0\abs{\bol E}^2+\frac{\abs{\bol B}^2}{\mu_0}\Big)d\,\bol r+\\
&+\int_\mathcal{V}\Big(\frac{\abs{\bol J_{ci}}^2}{\epsilon_0\omega^2_{pi}(\bf{r})}+\frac{\abs{\bol J_{ce}}^2}{\epsilon_0\omega^2_{pe}(\bf{r})}\Big) d\,\bol r, \quad\mathcal{V}\subset\mathbb{R}^3.
\end{split}
\end{equation}
The first integrand terms on the right hand side in Eq.\eqref{energy} correspond to the electromagnetic
field energy density, while the next two integrand terms are the kinetic energy density associated with the
electrons and ions in the plasma \cite{Bers_2016}.

\subsection{Discretization and encoding}\label{sec:2.2}
For simplicity we  will assume a $x-y$ uniform lattice with discretization step $\delta_x=\delta_y=\delta$ so for the configuration space  $\mathcal{V}=[x_0, x_0+L_x]\times[y_0, y_0+L_y]$ to be  comprised of $N_xN_y$ nodes separated by $\delta$ in each $x,y$ axes. As a result, in analogous way with Eq.\eqref{state}, the classical plasma state $\bol\psi$ in Eq.\eqref{state and conductivity} can be written as a $n_p=\log_2(N_xN_y)=n_{px}+n_{py}$ qubits pure state,
\begin{equation}\label{dicrete state}
\bol\psi(\bol r,t)\to\ket{\bol\psi(t)}=\sum_{p=0}^{2^{n_p} -1}\ket{\psi_q(t)}\otimes\ket{p}\in\mathcal{H}=\mathcal{H}_C\otimes\mathcal{H}_S,
\end{equation}
where the state $\ket{\psi_q}\in\mathcal{H}_C$  depends on the dimensionality of the plasma state $\bol\psi$ and it will be discussed later.

By defining the state $\ket{\psi_p(t)}=\ket{\psi_q(t)}\otimes\ket{p}$ we can express the momentum operator in the plasma Dirac equation in the discretized lattice space in each dimension using a  Euler difference scheme. For example in the in the $x$ direction for the forward difference,
\begin{equation}
\begin{split}\label{dicrete equation}
\frac{\ket{\psi_p(t+\Delta t)}-\ket{\psi_p(t)}}{\Delta t}+\textit{O}(\Delta t)&=\frac{c}{\Delta x}\Big(\ket{\psi_{p+1}(t)}\\
&-\ket{\psi_{p}(t)}\Big)+\textit{O}(\Delta x).
\end{split}
\end{equation}
Then, to first order accuracy $\Delta x \sim \delta$, $\Delta t= \Delta x/c\sim \delta$, Eq.\eqref{dicrete equation} reads,
\begin{equation}\label{discrete evolution}
\ket{\psi_p(t+\Delta t)}=\hat{S}\ket{\psi_{p}(t)}+\textit{O}(\delta^2).
\end{equation}
The $\hat{S}$ operator in Eq.\eqref{discrete evolution} is the streaming operator $\hat{S}\ket{p}=\ket{p+1}$. Therefore, the temporal evolution in Eq.\eqref{discrete evolution}, in the continuous limit $\delta\to0$ recovers to first order
\begin{equation}\label{first order}
i\pdv{\bol\psi(x,t)}{t}=-c\hat{p}_x\bol\psi(x,t)+\textit{O}(\delta).
\end{equation}
Applying forward $\hat{S}$ and backward $\hat{S}^\dagger$ steaming operations as well as alternating $\hat{C}$ and $\hat{C}^\dagger$ operations between different lattice cites  we aiming into a discretization scheme under which  the continuous evolution is recovered to second order $\textit{O}(\delta^2)$,  compared to Eq.\eqref{first order}, to ensure stability. The produced sequence is a QLA sequence that extends the DTQW evolution in Eq.\eqref{qw step evolution}.

\begin{figure}[h]
\centering
\includegraphics[width=0.8\linewidth]{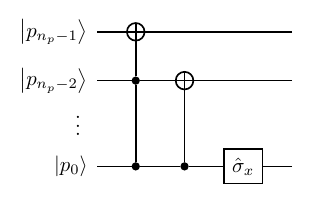}
\caption{Quantum gate implementation of streaming operator $\hat{S}$ in the $\ket{p}$ register. The least significant bit is the $p_{0}$. The recursive structure of the operation allows for a decomposition in $\textit{O}(n_p^2)$ singe-qubit and CNOT gates.}
\label{fig:1}
\end{figure}

The advantage of using the streaming operation in a quantum computer lies on its recursive structure \cite{Bao_2024} that allows for efficient implementation within $\textit{O}(n_p^2)$ elementary gates. Specifically, expressing $\ket{p}$ in its binary form $\ket{p_{n_{p}-1}p_{n_{p}-2}...p_{0}}$, 
the quantum respective circuit implementation of the unitary $\hat{S}$ operator is depicted in Fig.\ref{fig:1}.
Each of these $n_p$ in numbers multi-controlled CNOTs acting, at most, on $n_p$ qubits can be decomposed in $\textit{O}(n_p)$ elementary gates \cite{Barenco_1995} and therefore the total implementations scales as $\textit{O}(n_p^2)$.

What is left now is to encode the $\ket{\psi_q}$ state in the spin state to complement the walk process in the discretized lattice, allowing the action of coin operators to generate the evolution for Eq.\eqref{plasma dirac equation}.  In contrast with the Dirac quantum $4$-spinor the classical state $\bol\psi$ in Eq.\eqref{state and conductivity} is $12$-dimensional hence we need $n_q=4$ qubits in general to encode the spinor components 
in the spin space  $\mathcal{H}_C=\mathbb{C}^2\otimes\mathbb{C}^2\otimes\mathbb{C}^2\otimes\mathbb{C}^2$. However, the question that arises is now how to encode the $12$ classical amplitudes within the $16$ basis elements of $\mathcal{H}_C$ as pure state. We address this issue by considering the canonical form of a general $4$-qubit pure state $\ket{\psi}_4$. In that way, the  minimal number of local bases product
states in terms of which the state $\ket{\psi}_4$ can be written is twelve \cite{Acin_2001},
\begin{equation}\label{canonical form}
\begin{split}
\ket{\psi}_4&=a\ket{0000}+b\ket{0100}+c\ket{0101}+d\ket{0110}\\
&+e\ket{1000}+f\ket{1001}+g\ket{1010}+h\ket{1011}\\
&+i\ket{1100}+j\ket{1101}+k\ket{1110}+l\ket{1111}.
\end{split}
\end{equation}
Therefore we assign the $12$ states of the canonical $4$-qubit generalized Schmidt decomposition \eqref{canonical form} as the $\{\ket{q}\}$ register in respect of the components of the plasma state in Eq.\eqref{state and conductivity} as following:
\begin{equation}\label{amplitudes encoding}
    \begin{aligned}
    &\psi_0\leftrightarrow E_x\to \ket{q_0}\leftrightarrow\ket{0000}\\
    &\psi_1\leftrightarrow E_y\to \ket{q_1}\leftrightarrow\ket{0100}\\
     &\psi_2\leftrightarrow E_z\to \ket{q_2}\leftrightarrow\ket{0101}\\
     &\psi_3\leftrightarrow H_x\to \ket{q_3}\leftrightarrow\ket{0110}\\
     &\psi_4\leftrightarrow H_y\to \ket{q_4}\leftrightarrow\ket{1000}\\
     &\psi_5\leftrightarrow H_z\to \ket{q_5}\leftrightarrow\ket{1001}\\
     &\psi_6\leftrightarrow J_{cix}\to \ket{q_6}\leftrightarrow\ket{1010}\\
     &\psi_7\leftrightarrow J_{ciy}\to \ket{q_7}\leftrightarrow\ket{1011}\\
     &\psi_8\leftrightarrow J_{ciz}\to \ket{q_8}\leftrightarrow\ket{1100}\\
     &\psi_9\leftrightarrow J_{cey}\to \ket{q_9}\leftrightarrow\ket{1101}\\
     &\psi_{10}\leftrightarrow J_{cey}\to \ket{q_{10}}\leftrightarrow\ket{1110}\\
     &\psi_{11}\leftrightarrow J_{cez}\to \ket{q_{11}}\leftrightarrow\ket{1111}
    \end{aligned}
\end{equation}

Therefore, 
\begin{equation}
\ket{\psi_q}=\sum_{j=0}^{11}\psi_j\ket{q_j},
\end{equation}
and the total quantum encoded plasma state reads,
\begin{equation}
\ket{\bol\psi(t)}=\sum_{j,p}\psi_{j,p}(t)\ket{q_j}\otimes\ket{p}\in\mathcal{H}=\mathcal{H}_C\otimes\mathcal{H}_S.
\end{equation}

\section{The Quantum Algorithm}\label{sec:3}
The plasma Dirac Eq.\eqref{plasma dirac equation} is composed of a kinetic part $c\hat{\bol\gamma}_{em}\cdot\hat{\bol p}$ corresponding to the electromagnetic propagation in the vacuum space which has be treated before in \cite{Yepez_2002,Vahala_2020,Vahala_2023,Vahala2_2023,Vahala3_2023} using the inhomogeneous QW framework  where the unitary coin operators $\hat{C}$ depend on the discretization length $\delta$ \cite{Nzongani_2023}. In that way, we can retrieve the kinetic part to second order $\textit{O}(\delta^2)$ under a diffusion scheme $\Delta t= D\Delta x^2\sim \delta^2$  where $D=\textit{O}(1)$ is a diffusion coefficient. This is in sheer contrast with the the simple first order scheme presented in  Eqs.\eqref{dicrete equation}-\eqref{first order}.

On the other hand, the potential term $\Hat{V}(\bol r)$ in the matrix form in Eq.\eqref{potential} contributes only algebraically to Eq.\eqref{plasma dirac equation} and no streaming will be incorporated. To retrieve those terms we will need to include some external operators beyond the quantum walk process.

\subsection{The QLA sequence and the external operators}\label{sec:3.1}
We aim for an evolution sequence $t\to t+\Delta t$ of the form,
\begin{equation}\label{plasma sequence}
\ket{\bol\psi(t+\Delta t)}=\hat{V}_{pe}\hat{V}_{pi}\hat{V}_{ce}\hat{V}_{ci}\hat{\mathcal{U}}_{QLA}\ket{\bol\psi(t)},
\end{equation}
where $\hat{\mathcal{U}}_{QLA}$ is the unitary QLA sequence of streaming and coin operators related to the kinetic part whereas the $\hat{V}_{cj}$ and $\hat{V}_{pj}$ with $j=i,e$ are external unitary operators associated with the potential term $\hat{V}(\bol r)$.  Therefore, using the full sequence in Eq.\eqref{plasma sequence} we recover the plasma Dirac equation \eqref{plasma dirac equation} to second order $\textit{O}(\delta^2)$.

Fist we define the following coin operators, acting on each of the $x,y$ directions in the $12$-dimensional $\ket{\psi_q}\in\mathcal{H}_C$ spin state:
\begin{equation}\label{Cx}
\hat{C}_X=\begin{bmatrix}
\hat{C}_x &0\\
0& I_{6\times 6}
\end{bmatrix},
\end{equation}
with
\begin{equation}\label{CX}
\hat{C}_x=\begin{bmatrix}
 1&0&0&0&0&0\\
 0&\cos{\theta}&0&0&0&-\sin{\theta}\\
 0&0&\cos{\theta}&0&-\sin{\theta}&0\\
 0&0&0&1&0&0\\
 0&0&\sin{\theta}&0&\cos{\theta}&0\\
 0&\sin{\theta}&0&0&0&\cos{\theta}
\end{bmatrix},
\end{equation}
and
\begin{equation}\label{Cy}
\hat{C}_Y=\begin{bmatrix}
\hat{C}_y &0\\
0& I_{6\times 6}
\end{bmatrix},
\end{equation}
with
\begin{equation}\label{CY}
\quad \hat{C}_y=\begin{bmatrix}
 \cos{\theta}&0&0&0&0&\sin{\theta}\\
 0&1&0&0&0&0\\
 0&0&\cos{\theta}&\sin{\theta}&0&0\\
 0&0&-\sin{\theta}&\cos{\theta}&0&0\\
 0&0&0&0&1&0\\
 -\sin{\theta}&0&0&0&0&\cos{\theta}
\end{bmatrix},
\end{equation}
where the rotation angle $\theta$ is $\theta\sim c\delta/4$. The non trivial part of coin operators $\hat{C}_x$ and $\hat{C}_y$ are two-level unitary rotations acting locally on the $\ket{\bol E, \bol H}$ subspace of the overall $\ket{\psi_q}$ state. 

Then, the unitary QLA sequence for the kinetic part is $\hat{\mathcal{U}}_{QLA}=\hat{U}_Y\hat{U}_X$,
where $\hat{U}_X, \hat{U}_Y$ are the respective sequence of unitary coin-streaming operators in each direction,
\begin{widetext}
\begin{equation}\label{unitary x sequence}
\hat{U}_X=\hat{S}^{+x}_{25}\hat{C}^\dagger_X\hat{S}^{-x}_{25}\hat{C}_X\hat{S}^{-x}_{14}\hat{C}^\dagger_X\hat{S}^{+x}_{14}\hat{C}_X\hat{S}^{-x}_{25}\hat{C}_X\hat{S}^{+x}_{25}\hat{C}^\dagger_X\hat{S}^{+x}_{14}\hat{C}_X\hat{S}^{-x}_{14}\hat{C}^\dagger_X,
\end{equation}
\begin{equation}\label{unitary y sequence}
\hat{U}_Y=\hat{S}^{+y}_{25}\hat{C}^\dagger_Y\hat{S}^{-y}_{25}\hat{C}_Y\hat{S}^{-y}_{03}\hat{C}^\dagger_Y\hat{S}^{+y}_{03}\hat{C}_Y\hat{S}^{-y}_{25}\hat{C}_Y\hat{S}^{+y}_{25}\hat{C}^\dagger_Y\hat{S}^{+y}_{03}\hat{C}_Y\hat{S}^{-y}_{03}\hat{C}^\dagger_Y.
\end{equation}
\end{widetext}
In Eqs.\eqref{unitary x sequence} and \eqref{unitary y sequence} the streaming operator in each direction is defined similarly to Eq.\eqref{1d quantum walk} as
\begin{equation}
\hat{S}^{+x,y}_{ij}=(\ket{q_i}\bra{q_i}+\ket{q_j}\bra{q_j})\otimes \hat{S}^{+x,y}+\sum_{k-\{i,j\}}\ket{q_k}\bra{q_k}\otimes I_{2^p\times 2^p}.
\end{equation}
Since the streaming operator is unitary the walking process in the opposite direction is provided by $(\hat{S}^{+x,y}_{ij})^\dagger=\hat{S}^{-x,y}_{ij}$. The implementation of the streaming operator in the respective $\ket{p_x}$ and $\ket{p_y}$ registers follow that of Fig.\ref{fig:1} and therefore scales as $\textit{O}(n_{px}^2+n_{py}^2)$.

To recover the algebraic contribution from the potential in Eq.\eqref{potential} we decompose it into distinct contributions from the cyclotron and plasma frequency terms for each species,
structured terms as follows,
\begin{equation}\label{potential decomposition}
\hat V(\bol r)=\hat{D}_{\omega_{pi}}+\hat{D}_{\omega_{pe}}+\hat{D}_{\omega_{ci}}+\hat{D}_{\omega_{ce}},
\end{equation}
with an underlying  Pauli structure,
\begin{equation}
\hat{D}_{\omega_{pi}}=\frac{1}{2}\hat{\sigma}_y\otimes(I_{2\times2}+\hat{\sigma}_z)\otimes\omega_{pi},\label{Dpi}
\end{equation}
\begin{equation}
\hat{D}_{\omega_{pe}}=\frac{1}{2}(\hat{\sigma}_x\otimes\hat{\sigma}_y+\hat{\sigma}_y\otimes\hat{\sigma}_x)\otimes\omega_{pe}, \label{Dpe}
\end{equation}
\begin{equation}
\hat{D}_{\omega_{ci}}=\frac{1}{4}(I_{2\times2}-\hat{\sigma}_z)\otimes(I_{2\times2}+\hat{\sigma}_z)\otimes\omega_{ci}\hat{S}_z, \label{Dci}
\end{equation}
\begin{equation}
{D}_{\omega_{ce}}=\frac{1}{4}(I_{2\times2}-\hat{\sigma}_z)\otimes(I_{2\times2}-\hat{\sigma}_z)\otimes\omega_{ce}\hat{S}_z.\label{Dce}
\end{equation}
The $\hat\sigma$ matrices in Eqs.\eqref{Dpi}-\eqref{Dce} are the standard Pauli matrices.

As a result, to recover the non-differential terms associated with the plasma and magnetic inhomogeneity profiles, we have to define another set of unitary operators complementary to the QLA sequence. Specifically, for the diagonal cyclotron terms,
\begin{equation}\label{Vc matrices}
\hat{V}_{ci}=\begin{bmatrix}
I_{6\times6} &0_{6\times6}\\
0_{6\times6} &\hat{v}_{ci}
\end{bmatrix},\quad \hat{V}_{ce}=\begin{bmatrix}
I_{6\times6} &0_{6\times6}\\
0_{6\times6}&\hat{v}_{ce}
\end{bmatrix},
\end{equation}
with
\begin{widetext}
\begin{equation}\label{vc matrices}
\hat{v}_{ci}=\begin{bmatrix}
 \cos{\theta_{ci}}&-\sin{\theta_{ci}}&0&0&0&0\\
 \sin{\theta_{ci}}& \cos{\theta_{ci}}&0&0&0&0\\
 0&0&1&0&0&0\\
 0&0&0&1&0&0\\
 0&0&0&0&1&0\\
 0&0&0&0&0&1
\end{bmatrix},\quad \hat{v}_{ce}=\begin{bmatrix}
1&0&0&0&0&0\\
0&1&0&0&0&0\\
0&0&1&0&0&0\\
0&0&0&\cos{\theta_{ce}}&-\sin{\theta_{ce}}&0\\
0&0&0&\sin{\theta_{ce}}& \cos{\theta_{ce}}&0\\
0&0&0&0&0&1
\end{bmatrix}.
\end{equation}
\end{widetext}
The rotation angles now read $\theta_{ci,e}\sim \delta^2\omega_{ci,e}$. For simplicity we will assume homogeneous magnetic field so $\omega_{ci,e}=constant$. Similarly with the coin operators $\hat{C}_{X,Y}$ in Eqs.\eqref{Cx},\eqref{Cy} the cyclotron external potential operators are unitary rotation matrices acting on the locally in the spin register.

Moving on to the off-diagonal plasma inhomogeneity terms, we define $\hat{V}_{pi,e}$ as
\begin{equation}\label{Vpi}
\hat{V}_{pi}=\begin{bmatrix}
\cos{\theta_{pi}} &0 &-\sin{\theta_{pi}} &0\\
0 & I_{3\times3} &0 &0\\
\sin{\theta_{pi}}&0&\cos{\theta_{pi}}&0\\
0 &0&0&I_{3\times3}
\end{bmatrix},
\end{equation}
and
\begin{equation}\label{Vpe}
\hat{V}_{pe}=\begin{bmatrix}
\cos{\theta_{pi}} &0 &0&-\sin{\theta_{pe}}\\
0&I_{3\times3}&0&0\\
0&0&I_{3\times3}&0\\
\sin{\theta_{pe}}&0&0&\cos{\theta_{pe}}
\end{bmatrix}.
\end{equation}
The respective non-trivial elements in matrices \eqref{Vpi} and \eqref{Vpe} are diagonal $3\times3$ matrices with rotation angle $\theta_{pi}\sim \delta^2\omega_{pi}$. By considering an inhomogeneous plasma profile $\omega_{pi,e}=\omega_{pi,e}(\bol r)$ the external operators still posses a two-level rotational structure on the $4$ qubit spin space $\mathcal{H}_C$ but in contrast to the previous operators, their action changes in respect of the   $\{\ket{p}\}$ register since the rotation angle $\theta_{p}$ depends on the value of the plasma density profile $\omega_p$ in the lattice cite. Thus,
\begin{equation}\label{Vp action}
\hat{V}_p\ket{\bol\psi(t)}=\sum_{j=0}^{11}\sum_{p=0}^{2^{n_p}-1}\psi_{j,p}\hat{\mathcal{R}}_y(2\theta_p(p))\ket{q_j}\otimes\ket{p},
\end{equation}
with $ \theta_p(p)\sim \delta^2\omega_{p}(p)$. Take notice that in Eq.\eqref{Vp action} the superscripts in $\theta_p$ and $\omega_p$ denote quantities related to plasma density and they are not summation indices. For the same reason, we have also suppressed indices $i$ and $e$ for the respective plasma species. The operator $\hat{\mathcal{R}}_y$ denotes a two-level $\hat{R}_y$ rotation.

Finally, in the continuous limit $\delta\to0$  the sequence in Eq.\eqref{plasma sequence} recovers the Dirac representation of Maxwell system for the cold magnetized plasma \eqref{plasma dirac equation} to order $\textit{O}(\delta^2)$ under the diffusion ordering $\Delta t\sim\delta^2$. However, it must be highlighted that the sequence \eqref{plasma sequence} has been produced though a perturbation expansion in terms of $\delta$ to recover the kinetic and potential terms \eqref{Dpi}-\eqref{Dce} and not from Trotterization scheme i.e., separating  the exponential evolution operator into a product of the exponential parts of kinetic and potential operators respectively. This is evident since the respective operators in the sequence \eqref{plasma sequence} do not commute.

\subsection{Quantum circuit implementation}\label{sec:3.2}
In this section we provide explicit quantum circuit implementation of the participating operators in the evolution sequence  Eq.\eqref{plasma sequence} along with the respective gate cost, making the algorithm transparent and providing its implementation feasibility in actual hardware.

\begin{figure}[h]
\centering
\includegraphics[width=\linewidth]{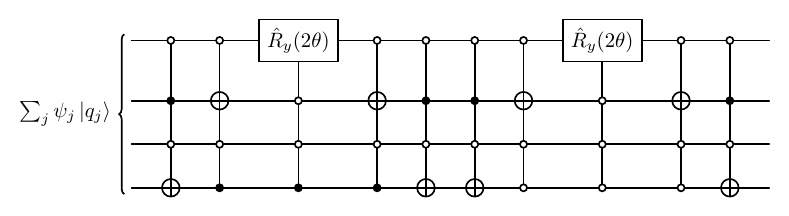}
    \caption{Quantum circuit implementation of the $\hat{C}_X$ operator in the $\{\ket{q_j}\}$ coin register. The spatial dependence has been suppressed for simplicity.}
    \label{fig:2}
\end{figure}

\begin{figure}[h]
    \centering
    \includegraphics[width=\linewidth]{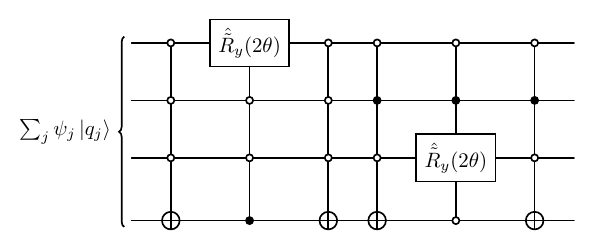}
    \caption{Quantum circuit implementation of the $\hat{C}_Y$ operator in the $\{\ket{q_j}\}$ coin register. The spatial dependence has been suppressed for simplicity.}
    \label{fig:3}
\end{figure}

The QLA $\hat{\mathcal{U}}_{QLA}$ is comprised by the streaming operators whose implementation has been provided in Fig.\ref{fig:1} and the coin operators which can be decomposed into two two-level $\hat{R}_y$ rotations acting locally in the coin space $\mathcal{H}_C$. Hence, the implementation of their action in the assigned canonical basis of Eq.\eqref{amplitudes encoding}, is depicted in Figs.\ref{fig:2} and \ref{fig:3}, where
\\
\begin{equation}\label{yrotations}
\hat{R}_y(\theta)=\begin{bmatrix}
\cos{\frac{\theta}{2}} & -\sin{\frac{\theta}{2}}\\
\sin{\frac{\theta}{2}} & \cos{\frac{\theta}{2}}
\end{bmatrix} \quad \text{and} \quad \hat{\Tilde{R}}_y=\hat{\sigma}_z\hat{R}_y\hat{\sigma}_z.
\end{equation}

Since each of the $\hat{C}_{X,Y}$ can be decomposed into two local two-level unitary operators acting on the $4$-qubit resister the implementations circuits in Figs.\ref{fig:2} and \ref{fig:3} can be decomposed into $\textit{O}(2\cdot4^2)$ single qubit and CNOT gates. Therefore, taking into consideration the respective sequences in Eqs.\eqref{unitary x sequence},\eqref{unitary y sequence} the overall implementation scaling into elementary quantum gates is $\textit{O}[16(n^2_{px}+n^2_{py}+32)]$. As a result, the qubit lattice algorithm simulates the kinetic part  $-c\hat{P}_{E,H}\otimes\hat{\bol\gamma}_{em}\cdot\hat{\bol p}$ in the plasma-Dirac representation Eq.\eqref{plasma dirac equation}, which reflects the  Maxwell equations in the vacuum \cite{Koukoutsis_2023}, within a number of elementary quantum gates (by dropping constant factors) $\textit{O}(n^2_{px}, n^2_{py})$. This gate complexity is similar by solving the respective free-part of the quantum Dirac or Maxwell equations using Quantum Fourier Transform (QFT).

\begin{figure}
    \centering
    \includegraphics[scale=1.0]{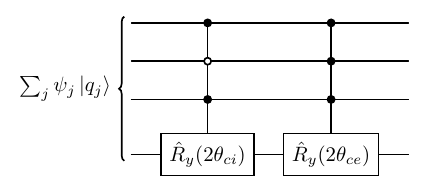}
    \caption{Quantum circuit implementation of the $\hat{V}_{ce}\hat{V}_{ci}$ product operator in the $\{\ket{q_j}\}$ coin register. The spatial dependence has been suppressed for simplicity.}
    \label{fig:4}
\end{figure}

The unitary potential operators associated with the cyclotron terms $\hat{V}_{ce,i}$ in Eqs.\eqref{Vc matrices} and \eqref{vc matrices} are two-level $\hat{R}_y$ unitary matrices acting only locally in $\mathcal{H}_C$ and admit an overall implementation as illustrated in Fig.\ref{fig:4}.
Based on this implementation, the decomposition of the product operator  $\hat{V}_{ce}\hat{V}_{ci}$ into elementary quantum gates scales as $\textit{O}(2\cdot 4^2)$.

\begin{figure}[h]
    \centering
    \includegraphics[width=\linewidth]{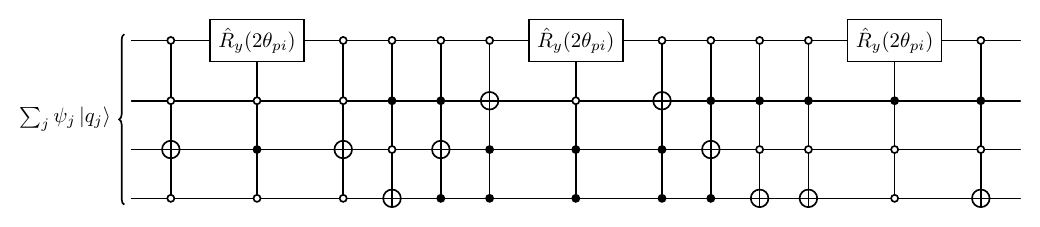}
    \caption{Quantum circuit implementation of the $\hat{V}_{pi}$ operator in the $\{\ket{q_j}\}$ coin register. The spatial dependence has been suppressed for simplicity.}
    \label{fig:5}
\end{figure}

\begin{figure}[h]
    \centering
    \includegraphics[width=\linewidth]{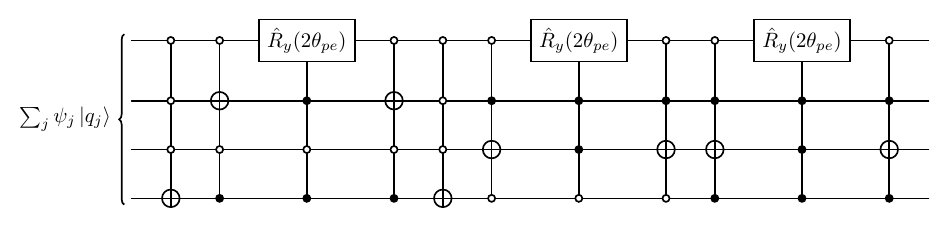}
    \caption{Quantum circuit implementation of the $\hat{V}_{pe}$ operator in the $\{\ket{q_j}\}$ coin register. The spatial dependence has been suppressed for simplicity.}
    \label{fig:6}
\end{figure}

Finally, each of the plasma density potential operators in Eqs.\eqref{Vpi},\eqref{Vpe} is a product of three two-level $\hat{R}_y$ rotations, resulting to a total implementation scaling of $\textit{O}(6\cdot 4^2)$ in the $\{\ket{q_j}\}$ register, according to the respective quantum circuits in Figs.\ref{fig:5} and \ref{fig:6}.

However, the action of the plasma density potential operators $\hat{V}_p$'s also depends on the $\{\ket{p}\}$ register according to Eq.\eqref{Vp action}. Therefore, the overall implementation scaling of the plasma density potential operators in the $\mathcal{H}_C\otimes\mathcal{H}_S$ space accounts for $\textit{O}(6\cdot 4^2\cdot 2^{n_p})$ elementary gates.

From the given construction, our algorithm is explicit without oracle operations and implicit operations and therefore can be implemented on actual quantum hardware.

\subsection{Complexity analysis and comparison with FDTD}\label{sec:3.3}
Following the discussion in Sec.\ref{sec:3.2}, the overall gate cost  $\mathcal{N}^q_{gate}(N)$ (number of elementary gates for quantum algorithm implementation) with $N$ being the number of lattice nodes for a time advancement $t\to t+\Delta t$ is $\mathcal{N}^q_{gate}(N)=N$. However, for physically relevant applications in electromagnetic scattering in magnetized plasmas from turbulent structures, the density profile of those is considered as a localized inhomogeneity imbued in the uniform background plasma density in the form of filaments or blobs \cite{Ram_2013,Ram_2016}. Therefore, in such scenario the potential operators $\hat{V}_p$ describing a localized inhomogeneity act,
\begin{equation}\label{poly action}
\hat{V}_p\ket{\bol\psi(t)}=\sum_{j=0}^{11}\sum_{p=0}^{poly(n_p)}\psi_{j,p}\hat{\mathcal{R}}_y(2\theta_p(p))\ket{q_j}\otimes\ket{p},
\end{equation}
with $\theta_p(p)\sim \delta^2\omega_{p}(p)$. Then, the respective gate cost is now $\mathcal{N}^q_{gate}(N)=poly(\log(N))$. In the homogeneous case, the only cost-significant gate is effectively the streaming operator, reducing the scaling to $\textit{O}(\log^2 N)$.

In comparison, the Finite Difference Time Domain (FDTD) algorithms that have been established as prominent tools in the computational studies of electromagnetic wave propagation and scattering \cite{Yee_1966,Schneider_1993,Lee_1999,Tsironis_2023} in complex media, use a staggered lattice in which evaluation of the electromagnetic  quantities $\bol E, \bol H$ in each lattice cite requires an additional interpolation in each lattice point for anisotropic  and inhomogeneous media such as plasmas \cite{Tsironis_2023}. As a result, the classical number of gates in the FDTD method for the same time advancement and number of lattice points is $\mathcal{N}^c_{gate}(N)=poly(N)$.

Following \cite{Papageorgiou_2013}, a measure for quantum advantage of the proposed quantum algorithm compared to the classical FDTD method is,
\begin{equation}\label{S1}
S_1(N)=\lim_{N\to\infty}\frac{\mathcal{N}^c_{gate}(N)}{\mathcal{N}^q_{gate}(N)}.
\end{equation}
Thus, the quantum algorithm exhibits an exponential quantum advantage compared to the FDTD method for localized inhomogeneities. The later speed-up reduces to polynomial for the general case of a global inhomogeneity profile reflecting fluctuations in a uniform background plasma density. 

Another measure to establish the quantum advantage is by considering the overall gate complexity for the total simulation time $T=N_t\Delta t$ in respect the desired accuracy $\varepsilon$. Then, the number of gates for simulation time $t\to t+T$ is $N_t\mathcal{N}_{gate}^{q,c}(N)$. The proposed quantum algorithm operates under a second order scheme, $\Delta t\sim\delta^2$ whereas the Courant–Friedrichs–Lewy (CFT) condition in FDTD connects the spatial resolution linearly with the time, $\Delta t\sim \delta$. Consequently, for the $2D$ case, $N_t=\textit{O}(N)$ for the QLA and $N_t=\textit{O}(\sqrt{N})$ for the FDTD. Therefore, the gate complexities for the total simulation time $\mathcal{N}^T_{gate}$ read,
\begin{equation}\label{total scalings}
\begin{aligned}
&\mathcal{N}^{q,T}_{gate}(N)= N^2,\quad \text{(general case)},\\
&\mathcal{N}^{q,T}_{gate}(N)= N poly(\log N),\quad \text{(localized inhomogeneity)},\\
&\mathcal{N}^{c,T}_{gate}(N)= \sqrt{N} poly(N),\quad \text{(FDTD)}.
\end{aligned}
\end{equation}
Equivalently, the gate complexities in Eq.\eqref{total scalings} in terms of the desired accuracy $\varepsilon$ read,
\begin{equation}\label{total accurancy}
\begin{aligned}
&\mathcal{N}^{q,T}_{gate}(\varepsilon)= T^4/\varepsilon^2,\quad \text{(general case)},\\
&\mathcal{N}^{q,T}_{gate}(\varepsilon)= (T^2/\varepsilon) poly\Big(\log \frac{T^2}{\varepsilon}\Big),\quad \text{(localized inhomogeneity)},\\
&\mathcal{N}^{c,T}_{gate}(\varepsilon)= (T^4/\varepsilon)^{1/3} poly\Big(\frac{T^{8/3}}{\varepsilon^{2/3}}\Big),\quad \text{(FDTD)}.
\end{aligned}
\end{equation}
The comparison between the respective ranges of gate complexities for the QLA and FDTD for $\kappa=1,2$ and $3$ dimensions, is presented in the Table \ref{table1}.
 \begin{table*} 
\caption{Gate scalings for the quantum QLA versus the  classical FDTD methods. The dimensionality of the problem is given by the parameter $\kappa=1,2,3$. \label{table1}}
 \begin{ruledtabular}
\begin{tabular}{l l l l }
            \textbf{Method} & \textbf{$\mathcal{N}_{gate}(N)$}& \textbf{$\mathcal{N}^T_{gate}(N)$} & \textbf{$\mathcal{N}^T_{gate}(\varepsilon)$} \\
             \toprule
            \textbf{FDTD}      & $\textit{O}[poly(N)]$ & $\textit{O}[N^{1/\kappa}poly(N)]$  & $\textit{O}[(T^2/\varepsilon)poly(T^2/\varepsilon)^\kappa]$ \\
            \textbf{QLA}       & $\textit{O}(\log^2 N)-\textit{O}(N)$ &  $\textit{O}(N^{2/\kappa}\log^2 N)-\textit{O}(N^{\frac{2+\kappa}{\kappa}})$   & $\textit{O}[(T^2/\varepsilon)\log^2(T^2/{\varepsilon})^{\kappa/2}]-\textit{O}[(T^2/\varepsilon)^{\frac{2+\kappa}{2}}]$ \\
        \end{tabular}
\end{ruledtabular}
 \end{table*}

Therefore, using Eq.\eqref{total accurancy} and Table \ref{table1} to evaluate the complexity comparison criterion \cite{Papageorgiou_2013},
\begin{equation}\label{S2}
S_2(\varepsilon)=\lim_{\varepsilon\to0}\frac
{\mathcal{N}^{c,T}_{gate}(\varepsilon)}{\mathcal{N}^{q,T}_{gate}(\varepsilon)},
\end{equation}
we obtain a strong  polynomial quantum speedup of the quantum algorithm compared to the FDTD for $\kappa\geq2$ and an exponential strong quantum speed up for the scattering off localized plasma inhomogeneities.

Consequently, the proposed qubit lattice algorithm not only possesses an explicit implementation structure but also exhibits, at worst case, polynomial advantage compared to the FDTD method for full-wave simulation of Maxwell simulations in cold inhomogeneous and magnetized plasmas.

\section{Generalizing to the dissipative case}\label{sec:4}
\subsection{The dissipative model}\label{sec:4.1}
Introducing the simplest form of dissipation requires the existence of a phenomenological collision frequency $\nu$ between the two species (ions-electrons) in plasma. Then, the frequency dependent  Stix permittivity matrix $\Tilde{\epsilon}_\nu(\omega)$ is \cite{Stix_1992,Bers_2016},
\begin{equation}\label{dissipativ Stix}
\Tilde{\epsilon}_\nu(\omega)=\begin{bmatrix}
S_\nu&-iD_\nu&0\\
iD_\nu&S_\nu&0\\
0&0&P_\nu
\end{bmatrix}
\end{equation}
with
\begin{align}\label{dissipative Stix elements}
S_\nu=&\epsilon_0\Big(1-\sum_{j=i,e}\frac{\omega^2_{pj}(\omega+i\nu)}{\omega(\omega+i\nu)-\omega_{cj}^2}\Big) \nonumber\\
D_\nu=&\epsilon_0\sum_{j=i,e}\frac{\omega_{cj}\omega^2_{pj}}{\omega[(\omega+i\nu)^2-\omega_{cj}^2]}\\
P_\nu=&\epsilon_0\Big(1-\sum_{j=i,e}\frac{\omega^2_{pj}}{\omega(\omega+i\nu)}\Big) \nonumber.
\end{align}
Obviously, now $\Tilde{\epsilon}_\nu(\omega)\neq \Tilde{\epsilon}_\nu^\dagger(\omega)$ since there is energy dissipation. For $\nu=0$ we recover the Hermitian (energy-preserving) counterpart $\Tilde{\epsilon}(\omega)$ in Eqs.\eqref{Stix},\eqref{Stix elements}.

In contrast with the conservative case in Eq.\eqref{susceptibility kernel}, the susceptibility kernel $\hat{K}_nu(\bol r, t)$ is characterized by both memory and dissipative effects, 
\begin{equation}\label{dissipative susceptibility}
\hat{K}_\nu(t)=\epsilon_0\sum_{j=i,e}\begin{bmatrix}
K_\nu^{(xx)}&K_\nu^{(xy)} &0\\
-K_\nu^{(xy)}&K_\nu^{(yy)}&0\\
0&0&K_\nu^{(zz)}
\end{bmatrix},
\end{equation}
with
\begin{equation}\label{dissipative susciptibility elements}
    \begin{aligned}
&K_\nu^{(xx)}=K_\nu^{(yy)}=\frac{\omega^2_{pj}}{\omega^2_{cj}+\nu^2}   \Big(e^{-\nu t}(\omega_{cj}\sin{\omega_{cj}t}-\nu\cos{\omega_{cj}}t)-\nu\Big)\\
&K_\nu^{(xy)}=\frac{\omega^2_{pj}}{\omega^2_{cj}+\nu^2}   \Big(e^{-\nu t}(\omega_{cj} \cos{\omega_{cj}}t+\nu\sin{\omega_{cj}t})+\omega_{cj}\Big)\\
&K_\nu^{(zz)}=\frac{\omega^2_{pj}}{\nu}(1-e^{-\nu t}). 
    \end{aligned}
\end{equation}
The total  conductivity current is now,
\begin{equation}\label{dissipative conductivity current}
\bol J_{\nu,c}=\int_0^t\pdv{\hat{K}_\nu(\bol r, t-\tau)}{t}\bol u(\bol r, \tau)d\,\tau,
\end{equation}
with 
\begin{equation}\label{relation between conservative and dissipative}
\pdv{\hat{K}_\nu}{t}=e^{-\nu t}\pdv{\hat{K}}{t}.
\end{equation}
The $\partial\hat{K}_\nu/\partial t$ term in Eq.\eqref{relation between conservative and dissipative} is the conservative counterpart, provided in \cite{Koukoutsis2_2023}. Based on the relation \eqref{relation between conservative and dissipative} and following the same procedure as in \cite{Koukoutsis2_2023} the resulted dissipative plasma-Dirac equation,
\begin{equation}
i\pdv{\bol\psi_\nu}{t}=\Big[-c\hat{P}_{E,B}\otimes\hat{\bol\gamma}_{em}\cdot\hat{\bol p} +\hat
V_\nu(\bol r)\Big]\bol\psi_\nu,
\end{equation}
now has the potential term $\hat{V}_\nu(\bol r)$ with an anti-Hermitian diagonal component,
\begin{equation}\label{dissipative plasma dirac}
\hat{V}_\nu(\bol r)=\begin{bmatrix}
0_{3\times3}&0_{3\times3}&-i\omega_{pi}&-i\omega_{pe}\\
0_{3\times3}&0_{3\times3}&0_{3\times3}&0_{3\times3}\\
i\omega_{pi}&0_{3\times3}&\omega_{ci}\hat{S}_z-i\nu&0_{3\times3}\\
i\omega_{pe}&0_{3\times3}&0_{3\times3}&\omega_{ce}\hat{S}_z-i\nu
\end{bmatrix}.
\end{equation}
In terms of the conservative generator of dynamics $\hat{D}=-c\hat{P}_{E,B}\otimes\hat{\bol\gamma}_{em}\cdot\hat{\bol p} +\hat V(\bol r)$, Eq.\eqref{dissipative plasma dirac} reads,
\begin{equation}\label{separated form}
i\pdv{\bol\psi_\nu}{t}=(\hat{D}-i\hat{D}_{diss})\bol\psi_\nu, \quad -i\hat{D}_{diss}=\hat{V}_\nu-\hat{V}.
\end{equation}
The $\hat{D}_{diss}$ operator in Eq.\eqref{separated form} is a diagonal Hermitian and positive definite matrix so for the $-i\hat{D}_{diss}$ to generate pure collisional dissipation.

\subsection{Post-selective time marching implementation procedure}\label{sec:4.2}
Treating the collisional plasma Dirac equation \eqref{separated form} in the context of unitary quantum computing can be accomplished though two distinct implementation roots which share a post-selective nature. 

The fist path consists of using a directly a first order Trotter product formula by separating the unitary from the non-unitary part. Then, according to Eq.\eqref{separated form} the Trotterized evolution reads
\begin{equation}\label{Trotterized evolution}
\ket{\bol\psi_{\nu}(t+\Delta t)}=e^{-i\Delta t\hat{D}}e^{-\Delta t\hat{D}_{diss}}\ket{\bol\psi_{\nu}(t)}+\textit{O}(\Delta t^2).
\end{equation}
The exponential non-unitary part in the Trotterized evolution Eq.\eqref{Trotterized evolution} can be easily evaluated as 
\begin{equation}\label{K matrix}
\hat{K}=e^{-\Delta t\hat{D}_{diss}}=diag(I_{6N\times 6N}, e^{-\nu \Delta t}I_{6N\times N}).
\end{equation}

Because we have established pure dissipation, the respective non-trivial diagonal elements of $\hat{K}$ matrix is can be written  as $e^{-\nu \Delta t}=\cos(\phi/2)$. As a result, we can decompose the non-unitary diagonal operator $\hat{K}$ in two unitary diagonal components,
\begin{equation}\label{Decomposition of K}
\hat{K}=\frac{\hat{K}_z+\hat{K}_z^\dagger}{2},
\end{equation}
with
\begin{equation}\label{Kz matrix}
    \hat{K}_z=\begin{bmatrix}
    I_{6N\times6N} &0\\
    0 & e^{-i\phi/2}I_{6N\times6N}
    \end{bmatrix}.
\end{equation}

Based on the unitary sum decomposition of the  the non-unitary matrix $\hat{K}$ in Eq.\eqref{Decomposition of K}, its implementation follows the Linear Combination of Unitaries (LCU) method \cite{Childs_2012} with the introduction of one ancillary qubit as follows. First we define the following unitary operators,
\begin{align}
\hat{U}_{prep}&:\ket{0}\to\frac{1}{\sqrt{2}}(\ket{0}+\ket{1})\label{preparation},\\
\hat{U}_{select}&=\ket{0}\bra{0}\otimes\hat{K}_{z}+\ket{1}\bra{1}\otimes\hat{K}_{z}^\dagger\label{selection},
\end{align}
where $\hat{U}_{prep}=\hat{H}$ is the Hadamard gate. The explicit form of the $\hat{U}_{select}$ operator,
\begin{equation}
\hat{U}_{select}=\begin{bmatrix}
\hat{K}_z &0\\
0 &\hat{K}^\dagger_z
\end{bmatrix},
\end{equation}
dictates that it is a diagonal unitary operator composed of $6N$ two-level $\hat{R}_z(\phi)$ gates and can be implemented within $\textit{O}[poly(n_p)]$ \cite{Hogg_1999}. In case we had different and inhomogeneous dissipation rates, then the implementation scaling would significantly increases to $\textit{O}(2^{n_p+1})$ \cite{Bullock_2004}.

\begin{figure}[h]
    \centering
    \includegraphics[width=\linewidth]{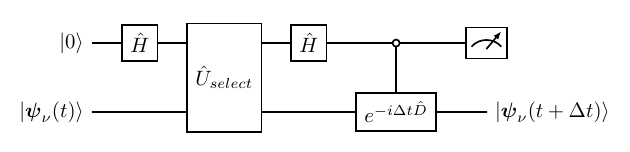}
    \caption{Quantum cirquit implementation of the non-unitary Trotterized evolution in Eq.\eqref{Trotterized evolution}. The implementation technique and scaling for the unitary operator $e^{-i\Delta t\hat{D}}$ has been detailed in Sec.\ref{sec:2}.}
    \label{fig:7}
\end{figure}

Then, by applying with the unitary operator 
\begin{equation}\label{dilated unitary}
\hat{\mathcal{U}}=(\hat{H}\otimes I_{12N\times 12N})\hat{U}_{select}(\hat{H}\otimes I_{12N\times 12N}).
\end{equation}
into the composite state $\ket{0}\otimes\ket{\bol\psi(t)}$, we can probabilistically implement the non-unitary $\hat{K}$ matrix,
\begin{equation}\label{output}
\hat{\mathcal{U}}(\ket{0}\otimes\ket{\bol\psi_{\nu}(t)})=\ket{0}\hat{K}\ket{\bol\psi_{\nu}(t)}+\frac{1}{2}\ket{1}(\hat{K}_{z}-\hat{K}_{z}^\dagger)\ket{\bol\psi_{\nu}(t)}.
\end{equation}
A unitary controlled operation in respect to the $0$-bit for the unitary part $e^{-i\Delta t\hat{D}}$, followed by a  measurement in the output state in Eq.\eqref{output} with respect to $\ket{0}$ state produces the non-unitary Trotterized evolution in Eq.\eqref{Trotterized evolution}. The respective quantum circuit implementation for the Trotterized time advancement $t\to t+\Delta t$ is depicted in Fig.\ref{fig:7}.

On a final note, the LCU dilation operation $\hat{\mathcal{U}}$ is a diagonalization of the one-qubit Sz.-Nagy dilation technique for the non unitary $\hat{K}$ operator \cite{Koukoutsis_2024,Schlimgen_2022},
\begin{equation}\label{Sznagy dilation}
\hat{\mathcal{U}}_{K}^{SN}=\begin{bmatrix}
\hat{K} & -\sqrt{I_{12N\times12N}-\hat{K}^2}\\
\sqrt{I_{12N\times12N}-\hat{K}^2} & \hat{K}
\end{bmatrix}.
\end{equation}
The Sz.-Nagy dilation matrix \eqref{Sznagy dilation} acts as a guaranteed block encoding of the non-unitary operator $\hat{K}$.

Let us now delve into the post-selective nature for a total simulation time of $T=N_t\Delta t$. For the infinitesimal time advancement $t\to t+\Delta t$ the success probability of the post selection is,
\begin{equation}\label{sucess probability}
p_{success}(\Delta t)=\mel{\bol\psi(t)}{\hat{K}^2}{\bol\psi(t)}.
\end{equation}
Expanding the $\hat{K}$ operator to first order in $\nu\Delta t$, Eq.\eqref{sucess probability} reads,
\begin{equation}\label{linearized probability}
p_{success}(\Delta t)\approx1-2\nu\Delta t\sum_{j=6}^{11}\sum_{p}\abs{\psi_{\nu, j,p}(t)}^2.
\end{equation}
In general, for $p_{success}(\Delta t)\sim \textit{O}(1)$ the Trotter time-step has to be selected as $\Delta t<<1/2\nu$.
The overall success probability $p_{sucess}(T)$, for implementing the normalized non-unitary evolution $\ket{\bol\psi_{\nu}(t)}\to\ket{\bol\psi_{\nu}(t+T)}/\norm{\psi_{\nu}(t+T)}$ with $\ket{\bol\psi_{\nu}(t+T)}=(e^{-i\Delta t \hat{D}}e^{-\Delta t\hat{D}_{diss}})^{N_t}\ket{\bol\psi_{\nu}(t)}$ to an error $\varepsilon$ \cite{Childs_2021} after $N_t$ repetitions of the quantum circuit in Fig.\ref{fig:7} with intermediate post-selections is,
\begin{widetext}
\begin{equation}\label{total sucess probability}
p_{success}(T)=\norm{\bol\psi_{\nu}(t+\Delta t)}^2\cdot\frac{\norm{\bol\psi_{\nu}(t+2\Delta t)}^2}{\norm{\bol\psi_{\nu}(t+\Delta t)}^2}\cdots\frac{\norm{\bol\psi_{\nu}(t+(N_t-1)\Delta t)}^2}{\norm{\bol\psi_{\nu}(t+(N_t-2)\Delta t)}^2}\cdot\frac{\norm{\bol\psi_{\nu}(t+T)}^2}{\norm{\bol\psi_{\nu}(t+(N_t-1)\Delta t)}^2}=\norm{\bol\psi_{\nu}(t+T)}^2.
\end{equation}
\end{widetext}
In Appendix \ref{Appendix}, it is demonstrated that the total success probability in Eq.\eqref{total sucess probability} of the proposed post-selective algorithm is non-vanishing, $p_{success}(T)\geq1/e$, in the limit $N_t\to\infty$. The latter has been firstly investigated by considering different time-scales and dissipation
strengths in \cite{Koukoutsis_2024} for simulating Maxwell equations in dissipative media and computationally demonstrated for simulating the advection-diffusion equation \cite{Over_2024}.

Consequently, the number of the required copies for obtaining the overall non-unitary evolution  from the normalized initial state $\ket{\bol\psi_{\nu}(t)}$ to the final normalized state is $1/p_{success}(T)\sim e$. Thus, the proposed post-selective scheme not only allows for an efficient implementation without requiring many copies of the initial state with parallel evolution but also avoids the need for amplitude amplification \cite{Brassard_2002} in the output state for a non-vanishing measurement. As a result, the quantum advantage established in the conservative case is retained in this post-selective protocol, as the implementation overhead scales at most by a multiplicative factor of $2e$.

Another implementation path would be the "QLAzation" of the dissipative equation \eqref{dissipative plasma dirac} resulting to a implementation sequence, similar with that of Eq.\eqref{plasma sequence},
\begin{equation}
\ket{\bol\psi(t+\Delta t)}=\hat{V}_{pe}\hat{V}_{pi}\hat{V}_{\nu,ce}\hat{V}_{\nu,ci}\hat{\mathcal{U}}_{QLA}\ket{\bol\psi(t)},
\end{equation}
where the operators $\hat{V}_{\nu,ce}, \hat{V}_{\nu,ci}$ are now non-unitary. Notice that since the dissipation is introduced though a diagonal form  it is expected that only the $\hat{V}_{ce}, \hat{V}_{ci}$ matrices will be affected. Once again, decomposing the non-unitary operators into a sum of unitary matrices enables the implementation using the LCU method.

\section{Conclusions}\label{sec:5}
Electromagnetic waves are ubiquitous in nature, playing a pivotal role in a wide range of real-world applications. In this paper, we explore the potential impact of quantum computing on the study of electromagnetic wave propagation and scattering in complex media by proposing a quantum algorithm to simulate Maxwell equations in magnetized plasmas. The scope of this paper aligns with efforts to leverage quantum computing as a powerful alternative to classical simulations in plasma physics and fusion research \cite{Dodin_2021,Amaro_2023,Joseph_2023,Koukoutsis2_2023}.

The main contributions of the paper are threefold. Firstly, the proposed qubit lattice algorithm for the energy-conserving case, features an explicit implementation structure suitable for testing on contemporary quantum hardware. More importantly, we have established a theoretical quantum speed-up over the widely used classical FDTD method for scattering studies in fusion plasmas. Finally, we develop a post-selective implementation procedure for non-unitary evolution in the presence of dissipation, modeled through a simple collisional mechanism, with an optimal, non-vanishing overall success probability. In that way, the number of the required copies for the proposed probabilistic quantum implementation to be successful is of order $\textit{O}(1)$. Consequently, the resource overhead remains a multiplicative factor of the conservative case, preserving the quantum advantage.

Our findings suggest that quantum computing has the potential to revolutionize the computational study of electromagnetic wave propagation and scattering in complex media. In the near future we will pursuit an actual implementation in quantum hardware to benchmark the theoretical performance of the quantum algorithm.

\section*{Acknowledgments}
This work has been carried out within the framework of the EUROfusion Consortium, funded by the European Union via the Euratom Research and Training Programme (Grant Agreement No 101052200 — EUROfusion). Views and opinions expressed are however those of the authors only and do not necessarily reflect those of the European Union or the European Commission. Neither the European Union nor the European Commission can be held responsible for them. E.K., K.H, and C.T., acknowledge valuable discussions with {\'O}. Amaro, L.I.I. Gamiz, A. Papadopoulos, I. Theodonis, Y. Kominis, P. Papagiannis and G. Fikioris.
A.K.R., G.V., M.S. and L.V. are  supported by the US Department of Energy under Grant Nos. DE-SC0021647, DE-FG02-91ER-54109, DE-SC0021651, DE-SC0021857 and DE-SC0021653.

\appendix
\section{Non-vanishing implementation probability}\label{Appendix}
By defining the normalized states,
\begin{equation}\label{eq1A}
\ket{\bol\phi(t+k\Delta t)}=\frac{\ket{\bol\psi_\nu(t+k\Delta t}}{\norm{\bol\psi_\nu(t+k\Delta t})},\quad k=0,1,...,N_t,
\end{equation}
with $\ket{\bol\phi(t)}= \ket{\bol\psi_\nu(t)},\,\,\abs{\ket{\bol\psi_\nu(t)}}=1$, the total success probability in Eq.\eqref{total sucess probability} reads,
\begin{equation}\label{eq2A}
p_{success}(T)=\prod_{k=0}^{N_t-1}\mel{\bol\phi({t+k\Delta t})}{\hat{K}^\dagger\hat{K}}{\bol\phi({t+k\Delta t})}.
\end{equation}
Substituting Eq.\eqref{K matrix} for the diagonal and  non-unitary operator $\hat{K}$ into Eq.\eqref{eq2A}
and taking advantage that normalization of the $\ket{\bol\phi}$ states, we obtain
\begin{equation}\label{eq3A}
p_{success}(T)=\prod_{k=0}^{N_t-1}\Big[1-(1-e^{-\beta})\sum_{j=6}^{11}\sum_p\abs{\phi_{j,p}(t+k\Delta t)}^2 \Big],
\end{equation}
with $ \beta=2\nu\Delta t$. Setting,
\begin{equation}\label{eq4A}
a_k=\sum_{j=6}^{11}\sum_p\abs{\phi_{j,p}(t+k\Delta t)}^2,\quad 0\leq a_k<1,
\end{equation}
the Eq.\eqref{eq3A} in the limit $N_t\to\infty$ is compactly written as an infinite product of the form,
\begin{equation}\label{eq5A}
\lim_{N_t\to\infty}p_{success}(T)=\prod_{k=0}^{\infty}\Big[1-(1-e^{-\beta})a_k \Big]
\end{equation}

The infinite product in Eq.\eqref{eq5A} converges to a non-zero positive number if and only if the following infinite sum converges \cite{Knopp_1990},
\begin{equation}\label{eq6A}
\sum_{k=0}^\infty a_k<\infty.
\end{equation}
Notice that in our case the $a_k$ in Eq.\eqref{eq4A} includes terms associated with the dissipative subspace defined by the $\hat{K}$ operator. In addition, the dissipation mechanism dictated by the form of $\hat{K}$ proposes that $a_k=a_{k-1}e^{-\beta}$. Thus,
\begin{equation}\label{eq7A}
a_k=a_0e^{-k\beta}.
\end{equation}
For more complex dissipative processes, instead of an exponential decay, a polynomial decay could be present $a_k=a_0 k^{-x}$. However, the infinite sum in Eq.\eqref{eq6A} converges, and therefore the probability is non-vanishing, only when $x>1$.

In the $\beta<<1$ limit (recall Eq.\eqref{linearized probability}) together with Eq.\eqref{eq7A}, the infinite product in Eq.\eqref{eq5A} takes the simple form,
\begin{equation}\label{eq8A}
P=\lim_{N_t\to\infty}p_{success}(T)=\prod_{k=0}^{\infty}(1-\beta a_0 e^{-k\beta})
\end{equation}
Taking the $\ln{P}$ and the approximation $\beta<<1$ we obtain,
\begin{equation}\label{eq9A}
\ln{P}=\sum_{k=0}^\infty\ln{(1-\beta a_0 e^{-k\beta})}\approx \beta a_0\sum_{k=0}^\infty e^{-k\beta}
\end{equation}
The infinite sum in Eq.\eqref{eq9A} is the limit of geometric series
\begin{equation}
\sum_{k=0}^\infty e^{-k\beta}=\frac{1}{1-e^{-\beta}}\approx \frac{1}{\beta},\quad \beta<<1.
\end{equation}
Finally,
\begin{equation}
\lim_{N_t\to\infty}p_{success}(T)=e^{\ln{P}}=e^{-a_0}\geq\frac{1}{e},
\end{equation}
since $0\leq a_0<1$.
\bibliography{ref}

\end{document}